\documentclass[aps,floats,prl,showpacs,twocolumn]{revtex4}

\usepackage{amsfonts,amsmath} 
\usepackage{bm} 
\usepackage{dcolumn}
\usepackage{epsfig} 
\usepackage{latexsym}

\begin{document}

\title{Low energy theory of disordered graphene}

\author{Alexander Altland}

\affiliation{Institut f{\"u}r Theoretische Physik, Universit{\"a}t zu
  K{\"o}ln, Z\"ulpicher Str. 77, 50937 K\"oln, Germany }

\date{\today}


\begin{abstract}
  At low values of external doping graphene displays a wealth of
  unconventional transport properties. Perhaps most strikingly, it
  supports a robust 'metallic' regime, with universal conductance of
  the order of the conductance quantum.  We here apply a combination
  of mean field and bosonization methods to explore the large scale
  transport properties of the system.  We find that, irrespective of
  the doping level, disordered graphene is subject to common
  mechanisms of Anderson localization. However, at low doping a number
  of renormalization mechanisms conspire to protect the conductivity
  of the system, to an extend that strong localization may not be seen
  even at temperatures much smaller than those underlying present
  experimental work.
\end{abstract}

\maketitle
The low energy physics of two--dimensional carbon, or
graphene\cite{novoselov05,zhang05}, is governed by the presence of two
generations of Dirac fermions. A conspiracy of these quasi--particles
with various channels of impurity scattering leads to a wealth of
intriguing transport phenomena which have attracted a lot of recent
attention (see Refs.~\cite{gusynin05a,ando05} for a list of early
references.) Prominent examples of the unorthodox conduction
properties of graphene monolayers include a quantum Hall effect with
unconventional plateaux quantization $\sigma_{xy}= (2n+1)e^2/h $, and
the experimental observation of a minimal universal conductivity
$\sigma_{xx}\sim e^2/h$ at low concentration of dopands, $E_F\sim
\tau^{-1}$~\cite{novoselov05,zhang05}. ($\tau^{-1}$ is the elastic
scattering rate and $E_F$ the Fermi energy, vanishing in the limit of
zero external doping.) While many of graphene's transport properties
have been successfully addressed theoretically, a number of important
questions remain open. Specifically, the nature of the low energy
regime $E_F\lesssim \tau^{-1}$ --- where the deviations from the
behavior of conventional metals are most pronounced --- is not fully
understood. Relying on calculations to lowest order in perturbation
theory in the disorder concentration~\cite{gusynin05}, the
experimental results have been interpreted~\cite{novoselov05} in terms
of some delocalized, genuinely metallic phase at the band
center. However, in the very low energy regime, where the conductivity
begins to level of to a disorder independent value, perturbation
theory breaks down and does not really help to elucidate the
physics of the system.

Building on a combination of mean field analysis and bosonized
perturbation around the clean Dirac fermion fixed point, $E_F=0$, we
here propose an effective low energy field theory of
system. Reflecting the large scale symmetries of the
problem~\cite{mccann06}, the relevant field theory at strong doping
$(E_F\tau \gg 1)$ is a standard nonlinear $\sigma$--model of either
orthogonal or unitary symmetry (depending on whether time reversal
invariance is broken or not.) The bare coupling constant of the theory
(cf. Eq.~(\ref{eq:3}) below) $\sigma_{xx}\sim E_F\tau\gg 1$ is large,
implying that localization corrections to the Drude conductivity are
comparatively weak.  Our main finding is that the weak doping regime,
$E_F\tau \ll 1$, too, is described by the nonlinear
$\sigma$--model. This result implies that at all doping levels
disordered graphene is subject to common mechanisms of Anderson
localization. Specifically, the smallness of the bare coupling
constant controlling the theory at $E_F\tau<1$ --- loosely to be
identified with the disorder independent 'minimal universal
conductivity' $\sigma_{xx}={\cal O}(1)$ of the system --- signals that
localization and fluctuation phenomena will be especially pronounced
at small doping. Below we will discuss a number of mechanisms which
may explain why these fluctuations have not yet been observed
experimentally. We conclude by briefly considering the Quantum Hall
physics of the system.

 In the absence of disorder, graphene's  two generations of low
energy Dirac
fermions are described by
the Hamiltonian~\cite{ando05} $\hat H_{\rm eff}= -iv\partial_x
\sigma_x \otimes \tau_3 -iv \partial_x \sigma_y$, where $v\sim ta$,
$a$ is the lattice spacing, the Pauli matrices $\sigma_i$ act in the
space of the two sublattices $A$ and $B$ of the system, and the Pauli
matrices $\tau_i$ act in two component nodal space. The
off--diagonality of the Hamiltonian in sublattice space implies
chirality, $[\hat H,\sigma_3]_+=0$.

In the presence of all possible channels of disorder scattering,
$\hat H$ generalizes to  (outer block structure in nodal space,
$2\times 2$ sub--blocks in sublattice space),
 \begin{equation}
  \label{eq:1}
 \hat H=\left[\begin{matrix}
V_A & -i v \partial+ V & W_A & W^{+-}\cr
-iv\bar{\partial}+ \bar V &V_B& W^{-+} & W_B\cr
\bar W_A & \bar W^{-+}& V_A&iv\bar{\partial}+ V' \cr
\bar W^{+-} & \bar W_B & iv\partial+ \bar V^{\prime}
&V_B \end{matrix}
\right],
\end{equation}
where $\partial = \partial_x + i \partial_y$, and $V_C$ and $W_C$,
respectively, describe the soft (intra--node) and the hard
(inter--node) scattering due to defect sites, next nearest neighbor
hopping, or disorder in adjacent substrates. Diagonal in sublattice
space, these scattering amplitudes violate the chiral symmetry of the
clean system. Similarly, the chiral symmetry preserving matrix elements
$V'$ and $W^{+-}, W^{-+}$ represent the intra-- and inter--node
amplitudes, respectively, of defective bond hopping.

Referring for a much more comprehensve discussion,
to Ref.~\cite{mccann06}, we briefly comment on the symmetries of the problem: (i) if all hopping matrix elements of the
original problem are real (complex), the low energy Hamiltonian does
(not) obey the symmetry relation $\hat H = \tau_1 \hat H^T \tau_1$,
thus falling into the Wigner--Dyson orthogonal (unitary) symmetry
class.  However, (ii) depending on the microscopic realization of the
disorder, some of the amplitudes entering the bare Hamiltonian may be
much smaller (even negligibly small at the length scales probed in
current experiment) than others.  If this happens the approximate
realization of symmetries beyond Wigner--Dyson
orthogonal/unitary may result in various transport anomalies,
including weak antilocalization phenomena at intermediate length
scales~\cite{suzuura02,mccann06,nomura06}. However, not protected by
rigid exclusion principles, these structures are transient in nature
and will not affect the physics at the largest distance scales. As we
are primarily interested in the latter, we will assume the presence of
all disorder channels throughout.

Transport properties of the system are described by disorder averaged
products of retarded and advanced Green functions $\langle (E_F-\hat
H+i0)_{ {\bf x}_1, {\bf x}_2}^{-1} (E_F-\hat H-i0)_{{\bf x}_3, {\bf
    x}_4}^{-1}\dots \rangle_{\rm dis}$.  To compute such correlation
functions we consider the replica coherent state field
integral (generalization to a supersymmetry
formulation is straightforward)
\begin{equation}
  \label{eq:2}
  \mathcal{Z}= \int D(\bar\psi,\psi) \,\left \langle \exp\left(i \int d^2 x\, \bar \psi(i\delta
\sigma_3^{\rm ar} E_F- \hat H )\psi\right)\right\rangle_{\rm dis},
\end{equation}
where $\psi=\{\psi_{s,a,\nu}({\bf x})\}$ is a $(2\times R\times
4)$--component Grassmann field, $s=1,2$ is a two
component index distinguishing between retarded and advanced indices,
$a=1,\dots,R$ is a replica index, $\nu=1,\dots,4$ enumerates
sublattice and nodal sectors and $\sigma_3^{\rm ar}\equiv \delta_{ss'}
(-)^{s+1}$ is a Pauli matrix causing infinitesimal symmetry breaking
in advanced/retarded space.

Beginning with the {\it strong doping regime}, $E_F\tau\gg 1$, we next
outline the construction of the low energy field theory describing the
large scale behavior of the system (i.e., at length scales larger than
the elastic mean free path, $l$.)  The construction of the theory at
large doping follows standard procedures
(cf. Refs.~\cite{efetov97,altland06} for review) and we restrict
ourselves to a brief recapitulation of the central construction steps:
the action in (\ref{eq:2}) is invariant under uniform unitary
rotations $\psi\to U\psi$, where $U\in {\rm U}(2R)$. Averaging over
all scattering channels in (\ref{eq:1}) generates a mean field fermion
self energy $i\delta \sigma_3^{\rm ar} \to i \Sigma \sigma_3^{\rm
  ar}$, breaking the symmetry down to ${\rm U}(R)\times {\rm
  U}(R)$. Along with this spontaneous symmetry breaking, Goldstone
modes $Q\in {\rm U}(2R)/{\rm U}(R) \times {\rm U}(R)$ appear. The 
effective action $S[Q]$ controlling the low energy partition
function ${\cal Z}_{\rm eff}\equiv \int DQ \,\exp(-S[Q])$ of these
modes is fixed by noting that the field space admits a (unique) parity
invariant second order derivative operator, $-\int d^2x \,{\rm
  tr\,}(\partial_\mu Q\partial_\mu Q)$. Gauge arguments (but also the
microscopic construction) state that the bare coupling of this
operator is $(1/8)\times$ the dimensionless Drude longitudinal
conductivity, $\sigma_{xx}$, i.e.
\begin{equation}
  \label{eq:3}
 S[Q] = -{\sigma_{xx}\over 8}\int d^2 x \,{\rm
  tr\,}(\partial_\mu Q\partial_\mu Q),
\end{equation}
which is the standard nonlinear $\sigma$--model action of time
reversal invariance broken weakly disordered metallic systems. (In the
time reversal invariant case, $Q\in {\rm Sp}(4R)/{\rm Sp}(2R)\times
{\rm Sp}(2R)$ and $\sigma_{xx}/8\to \sigma_{xx}/16$.)  Specifically,
graphene's Drude conductivity is given by~\cite{gusynin05}
$\sigma_{xx}\stackrel{E_F\tau \gg 1}= 2\tau E_F$. Referring for an in
depth discussion of the localization properties of the theory
(\ref{eq:3}) to Ref.~\cite{efetov97}, we note that at intermediate
length scales $L>l$ this bare value will be modified by weak
localization corrections (for a comprehensive discussion of weak
localization effects in graphene at $E_F\tau>1$ and different channels
of disorder scattering, see Ref.~\cite{mccann06}), before strong
Anderson localization sets in at exponentially large scales $L\sim
\xi\equiv l \exp({\rm const.}\times \sigma^2_{xx})$ (unitary symmetry)
or $\xi\equiv l \exp({\rm const.}\times \sigma_{xx})$ (orthogonal
symmetry.) However, before discussing the phenomenology of the system
any further, let us explore what happens in the strong disorder, or
{\it weak doping regime}, $E_F\tau < 1$.

While the mean--field type derivation of the action (\ref{eq:3})
essentially relied on the presumed largeness of the parameter
$E_F\tau$, changes in the bare disorder concentration are not expected
to change the basic symmetries of the system. One may, thus, suspect
that the low energy theory retains its integrity as we push on into
the strong disorder regime $E_F\tau <1$.
In the following we will apply bosonization techniques (similar to
those previously used in the context of the $d$--wave
superconductor~\cite{nersesyan94,asz02}) to confirm that this
conjecture is correct. Readers not interested in the technicalities of
this construction may directly advance to the discussion two
paragraphs further down.

Our starting point is Wittens seminal result~\cite{witten84} according
to which a system of $N$ free Dirac fermions may be equivalently
represented in terms of a level 1 Wess--Zumino--Witten (WZW) model
with target manifold $U(N)$: $ \int d^2x \left(\bar \psi_+ \partial
  \psi_+ + \bar \psi_- \bar \partial \psi_-\right) \leftrightarrow
S[M]$, where $\psi_\pm$ are $N$--component Grassmann fields and $M\in
{\rm U}(2R)$ is an $N\times N$ unitary matrix field with action
$S[M]=S_0[M]+ {i\over 12\pi} \Gamma[M]$. Here, $S_0[M]={1\over
  8\pi}\int d^2x\,{\rm tr\,}(\partial_\mu M \partial_\mu M^{-1})$ and
$\Gamma[M]$ is the topological WZW--action whose detailed structure
will not be of relevance throughout~\cite{fn_WZW}. Specifically, the
replica theory of clean graphene with its two generations ($K$ and
$K'$) of $2R$ Dirac fermions is represented in terms of two matrix
fields $M,M'\in {\rm U}(2R)$, with decoupled action
$S[M,M']=S[M]+S'[M']$. Here $S[M]=S_0[M]+ {i\over 12\pi} \Gamma[M]$
and $S'[M']=S_0[M']- {i\over 12\pi} \Gamma[M']$ where the relative
sign in front of the WZW--actions signals the different orientation
(or parity) of the two low energy islands. The role of disorder
scattering may now be explored by averaging the exponentiated impurity
matrix elements in the prototypical action (\ref{eq:2}) over the
disorder distribution functions (here assumed to be Gaussian
correlated.) This results in the appearance of four--fermion
operators, which may be subsequently be represented in bosonic
language~\cite{witten84}. Exemplifying this procedure on the
inter--node chiral symmetry preserving scattering amplitudes, we find
\begin{widetext}
$$
\left\langle e^{-i   \int d^2x\, \left[\bar \psi_+ W^{+-}
  \psi_+' + \bar \psi_-' \bar W^{+-} \psi_-\right]}\right\rangle_{\rm
dis.}=e^{ \gamma^{KK'}_{AB} \int d^2 x\,{\rm tr}(\psi_-\bar\psi_+
\, \psi_+'\bar\psi_-')}\to e^{  \gamma^{KK'}_{AB} a^{-2} \int d^2 x\,{\rm tr}(M 
 M^{\prime -1})},
$$
\end{widetext}
where the constant $\gamma^{KK'}_{AB}$ is a measure of strength of the
disorder, $a$ is the small--distance cutoff of the theory (the lattice
spacing, say), and the correspondence $\psi_{-a} \bar
\psi_{+b}\leftrightarrow a^{-1} M_{ab}$ was used. In a similar
manner, the scattering channels $V_{A,B}$ generates operators
$-\gamma_{A} a^{-2} \int d^2x \left[{\,\rm tr\,}(M+M^{\prime
    -1})\right]^{-2}$. 

To understand the consequences of the disorder generated modification
of the theory, we note that the two operators above are examples of
marginally relevant perturbations at the clean Dirac fermion fixed
point, $E_F=0$~\cite{ludwig94,asz02}: For initially weak disorder, the constants $\gamma_A$
and $\gamma_{AB}^{KK'}$  grow to values of ${\cal O}(1)$ only at
exponentially large length
scales $l\sim a \exp({\rm const.}\times v_F^2/\gamma_0)$, where
$\gamma_0$ represents the bare values of the coupling. The crossover
scale, $l$, {\it defines} the elastic mean free path of the theory. At
larger scales, the inter--node scattering operator ${\rm tr}(MM')$
enforces locking $M=M'$ of the formerly independent nodes while node
and sublattice diagonal scattering, $[{\rm tr}(M+M')]^{-1}$, implies a
field reduction $M\to Q\in {\rm U}(2R)/{\rm U}(R)\times {\rm U}(R)$
down to the manifold discussed earlier in the context of the weak
coupling regime. Substituting the reduced fields into the action
$S[M,M']\to S[Q,Q]$ (For the legitimacy of naive massless field
substitution within a strong coupling context, see
Ref.~\cite{affleck87}) we obtain Eq. (\ref{eq:3}) at
$\sigma_{xx}=4/\pi$. (For time reversal invariant systems, a similar
construction leads to the orthogonal variant of the theory.)

We conclude that the theory describing electronic transport in
graphene at low doping, $E_F\tau\ll 1$, is a nonlinear $\sigma$--model
with disorder independent bare coupling constant $4/ \pi$.  However,
unlike with the weak coupling situation discussed above, (a), this
description becomes valid only at length scales $l\sim a \exp({\rm
  const.}\times v_F^2/\gamma_0)$, exponentially large in the bare
inverse disorder concentration. At smaller length scales, the dynamics
is essentially ballistic. (Within the framework of a large $N$ mean
field approach, this strong disorder sensitivity of the Dirac fermion
scattering mean free path, and the relevancy of the $\sigma$--model at
larger scales was first observed in Ref.~\cite{fradkin86}.) In
contrast, the mean free path $l\sim \gamma^{-1}$ at strong doping
$E_F\tau \gg1$ depends much weaker on the disorder concentration and
the crossover to a multiple scattering regime occurs at smaller length
scales. Second, (b) the smallness of the coupling constant $4/\pi$
implies that the bare value of the conductivity will be subject to
{\it strong} renormalization. More precisely, large fluctuations of
the $Q$--fields (describing the proliferation of the particle--hole
modes causing Anderson localization) will lead to an exponential decay
of correlation functions at length scales $\xi={\cal N} l$, where
${\cal N}$ is a numerical factor. Accordingly, the conductivity of an
infinite graphene sheet $\sigma_{xx}(T=0)=0$, while the conductance of
a sample with extension $L\gg l$ will be exponentially small in
$\exp(-{\rm const.}\times L/\xi)$. However, as first pointed out
in~\cite{fradkin86}, the value ${\rm min}\,(L,l_\phi)$ beyond which
the conductivity is strongly diminished may be orders of magnitude
larger than $l$: naive application of a 1--loop RG to the strong
coupling problem (\ref{eq:3}) suggests that for systems with broken
time reversal invariance, localization sets in  at scales $\xi
\sim e^{16} l$.  In time reversal invariant systems $\xi \sim
e^{4} l$, still about two orders of magnitude larger than $l$.

How can these results be reconciled with the experimental observation
of a universal value $\sigma_{xx,{\rm exp}}\simeq 4$, significantly
larger than even the bare (Drude) conductivity of the {\it clean}
Dirac system? At this stage, the numerical discrepancy to the Drude
value is not fully understood (see, however, Ref.~\cite{nomura06}.) As
for the general robustness of the minimal conductivity, (i) elastic
mean free path renormalization at small doping and (ii) the largeness
of the sigma model correlation length even at strong coupling conspire
to render the length scale beyond which strong localization is seen
large. Finally, (iii), we repeat that the above discussion applied to
a regime where all disorder scattering channels are of comparable
strength. On the other hand, the experimentally observed linear
dependence of the conductivity on the carrier density
suggests~\cite{nomura06} dominance of the soft (node diagonal)
potential scattering, $V_A=V_B$ channel. If only this channel were
operational, the relevant field theory would be two independent copies
(the nodes) of sigma models of symplectic symmetry (technically,
Witten's theory subject to the constrait $M=Q$, $M'=Q'$), each
generating weak {\it anti}localization. However, beyond a certain
crossover scale, internode scattering will become effectual, the field
theory gets locked to a single copy of a theory of orthogonal symmetry
and standard Anderson localization emerges. We thus conclude that
graphene is subject to ordinary mechanisms of Anderson
localization. Although, a conspiracy of points (a-d) above may imply
that the observed metallic regime is extraordinarily robust, weak
localization precursors of strong localization will likely be observed
at temperatures lower than the present $10K$.

Slightly modifying a line of arguments originally due to
Pruisken\cite{pruisken84}, we finally touch upon the quantum Hall
regime. On very general grounds, i.e. regardless of whether a magnetic
fields is present or not, the theory contains a first order derivative
operator, $S^{(1)}[T]= - {\pi \over 2}\int d^2x \,\langle j_\mu ({\bf
  x}) \rho(E_F) \rangle\, {\rm tr}(\sigma_3^{\rm ar}
T^{-1} \partial_\mu T)$, where $j_\mu({\bf x})$ is the local current
density operator, $\rho(E_F)$ the (disorder averaged) density of
states at the Fermi energy and $\langle \dots \rangle_q$ the trace
over single particle Hilbert space. While the bulk expectation value
$\langle j_\mu \rho\rangle_q$ generally vanishes, the presence of
quantum Hall edges couples $S^{(1)}$ to the theory. Substitution of
the edge state system of graphene\cite{neto06,brey06} into
$S^{(1)}[T]$ obtains the boundary operator $S^{(1)}[T]= - \nu\oint ds
\,\langle {\rm tr}(\sigma_3^{\rm ar} T^{-1} \partial_s T)$, where $s$
parameterizes the system boundary and $\nu=2n+1$ is the number of
Landau levels below the Fermi energy. Applying Stokes theorem to
convert $S^{(1)}$ to a bulk integral, adding the background action
(\ref{eq:3}), and using that $\nu=\sigma_{xy}/2$ for Fermi energies
intermediate between Landau levels, we obtain
\begin{equation}
  \label{eq:4}
 S[Q] = -{1\over 8}\int d^2 x \,{\rm tr}\left[\sigma_{xx} \partial_\mu Q\partial_\mu Q- \sigma_{xy} \epsilon_{\mu\nu}Q\partial_\nu Q \partial_\mu Q\right].
\end{equation}
For high Landau levels, (\ref{eq:4}) is but a variant of Pruiskens
quantum Hall action; the bare longitudinal conductivity $\sigma_{xx}$
is large and the structure of graphene's edge spectrum merely enters
through the odd--integer quantization of the Hall conductivity. With
the lowest Landau level (LLL), $\nu=1$, the situation is more
interesting. Referring for the technicalities of bosonization in the
presence of magnetic fields and system boundaries to
Refs.~\cite{asz02,delphenich97}, we note that the applicability of the
action (\ref{eq:4}) to the LLL must be taken with a grain of salt: for
one thing, the bosonization approach {\it perturbs} around the clean,
non--magnetic limit of the theory. Quantities such as the disorder
averaged broadened density of states of the LLL are likely
inaccessible in terms of (renormalized) perturbation theory around
that point.  Second, and unlike with the situation in ordinary metals,
the conductivity tensor of the LLL $(\sigma_{xx},\sigma_{xy})=2\times
(2/\pi, 1)$ is numerically close to the prospected value of the
quantum Hall {\it transition} point
$(\sigma_{xx}^\ast,\sigma_{xy}^\ast)$. This is important inasmuch as
the QHE transition is arguably~\cite{zirnbauer99,bhaseen00} not
described by Pruiskens theory. In practice, this may imply that 
strong field fluctuations (to be expected on account of the smallness
of the coupling constants) readily compromise the integrity of the LLL
action~(\ref{eq:4}). At any rate, the identity of the proper theory of
the quantum Hall transition remains as unknown as it is in normal
metals.

Summarizing, we have constructed the low energy theory of disordered
graphene at large length scales. In its low doping regime,
$E_F\tau<1$, the system is described by a nonlinear $\sigma$--model at
strong coupling. Derived by bosonization methods, this model
consistently matches the weak coupling mean field approach, valid at
large doping $E_F\tau\gg 1$. Our main finding is that the 'universal
minimal conductivity' in graphene will be subject to conventional
mechanisms of quantum interference at temperatures lower than those
underlying current experiment. However, depending on the symmetries
actually realized in the system, the temperatures below  which the
conductance is {\it strongly} diminished by localization may be extremely low.

Discussions with R. Egger, E. McCann, B. Simons, V.Falko and M. Zirnbauer are
gratefully acknowledged. I thank E. Fradkin for pointing out to me the
profound applied relevance mean free path renormalization likely has to
the physics of the low doping regime.  This work was supported by
Transregio SFB 12 of the Deutsche Forschungsgemeinschaft. Note added:
after completion of this work, I became aware of the closely related manuscript
I.L.~Aleiner and K.B.~Efetov, cond-mat/0607200.


\end{document}